\pdfoutput=1
\documentclass[12pt]{iopart}
\usepackage{graphicx}

\begin{document}

\newcommand{\pasj}{PASJ}
\newcommand{\aap}{A\&A}
\newcommand{\mnras}{MNRAS}
\newcommand{\apj}{ApJ}
\newcommand{\apjl}{ApJ}
\newcommand{\app}{APh}
\newcommand{\memsai}{MmSAIt}
\newcommand{\ssr}{Space Science Reviews}
\newcommand{\nat}{Nature}
\newcommand{\aapr}{Astronomy \& Astrophysics Rev.}
\newcommand{\araa}{Annual Rev. of Astronomy \& Astrophysics}
\newcommand{\apss}{Astrophysics and Space Science}
\newcommand{\jcap}{Journal of Cosmology and Astroparticle Physics}
\newcommand{\prd}{Physical Review D}
\newcommand{\physrep}{Physics Reports}
\newcommand{\na}{New Astronomy}
\newcommand{\aj}{Astronomical Journal}

\title[Turbulent acceleration in galaxy clusters]{The challenge of turbulent 
acceleration of relativistic particles in the intra-cluster medium}

\author{Gianfranco Brunetti}

\address{INAF- Osservatorio di Radioastronomia, via P. Gobetti 101,
I-40129, Bologna, Italy}

\ead{brunetti@ira.inaf.it}
\begin{abstract}
Acceleration of cosmic-ray electrons (CRe) in the intra-cluster-medium (ICM) 
is probed by radio observations that
detect diffuse, Mpc-scale, synchrotron sources in a fraction of galaxy clusters.
{\it Giant radio halos} are the most spectacular manifestations of
non-thermal activity in the ICM and are currently
explained assuming that turbulence driven during massive cluster-cluster
mergers reaccelerates CRe at several GeV.
This scenario implies a hierarchy of complex mechanisms in the ICM that drain 
energy from large-scales into electromagnetic fluctuations in the
plasma and 
collisionless mechanisms of particle acceleration at much smaller scales.
In this paper we focus on the physics of
acceleration by compressible turbulence. The spectrum and damping
mechanisms of the electromagnetic fluctuations, and the mean-free-path (mfp)
of CRe are the most
relevant ingredients that determine the efficiency of acceleration. 
These ingredients in the ICM are however
poorly known and we show
that calculations of turbulent acceleration are also sensitive to
these uncertainties.
On the other hand this fact implies 
that the non-thermal properties of galaxy
clusters probe the complex microphysics 
and the {\it weakly collisional} nature of the ICM.
\end{abstract}


\section{Introduction}

Clusters of galaxies are the largest virialized structures in
the universe \cite{borgani}.
They form via a hierarchical sequence of mergers and accretion of smaller 
systems driven by dark matter that dominates the gravitational field.
The majority of baryonic matter in clusters is in the form 
of a hot (T $\sim 10^8$ K) and tenuous ($N_{gas} \sim 10^{-1} - 10^{-4}$ 
cm$^{-3}$) gas, the intra-cluster-medium (ICM).

The ICM is a unique environment for plasma physics due to the
combination of its weakly collisional nature and high plasma beta.
It also provides situations for CR acceleration and dynamics
that differ significantly from those in other astrophysical
environments.
First of all,
the bulk of CRs are confined and accumulated in the very large
volumes of galaxy clusters for Hubble time
\cite{voelk96, berezinsky97,blasi07,brunettijones14}.
Second, the life-time of CRs in the ICM is
very long as CRs diffuse in a dilute high-beta plasma
\cite{sarazin99,brunettijones14}.
Under these conditions particle acceleration
mechanisms that are notoriously not very efficient, 
such as stochastic Fermi-II mechanisms, become important because they 
can act for very long times-scales and on very large volumes.
Another point is that the combination of CRs confinement and long 
life-time with the complex dynamics of the ICM implies that CRs 
can experience phases of cooling and (re)acceleration by different
mechanisms. This results in a complex energy and spatial distribution 
of the CRs in clusters\cite{brunettijones14,pinzke15}.

Clearly galaxy clusters contain several {\it discrete} sources of CRs, 
including galaxies and AGNs. However there is much more.
Indeed radio observations show cluster-scale diffuse synchrotron
sources in a fraction of galaxy clusters, the most prominent ones are {\it
giant radio halos} that cover Mpc$^3$ volumes \cite{feretti12}.
These emissions are not directly associated
with discrete sources and require {\it in situ}
mechanisms of particle acceleration in the ICM \cite{brunettijones14}.
{\it Radio halos} are observed in cluster mergers
\cite{feretti12,cassano10,cassano13,kale15}, suggesting that
a fraction of the kinetic energy of large-scale motions can be 
channelled into electromagnetic fluctuations and particle
acceleration at smaller scales in the ICM.
In fact this point is telling us that a complex 
hierarchy of processes are active in the ICM at different scales, and 
offers a possibility to constrain fundamental aspects of 
the microphysics of the ICM.

\noindent
The synchrotron spectra of {\it radio halos} are steep with
spectral slopes measured at GHz frequencies $\alpha \sim 1-2$
(flux $\propto \nu^{-\alpha}$) suggesting that the underlying
acceleration mechanisms are not very efficient \cite{brunettijones14}.
A popular scenario for the origin of {\it radio halos} is based on the
possibility that turbulence generated during cluster-cluster mergers
(re)accelerates seeds electrons distributed on Mpc$^3$-volumes
to energies $\geq$ few GeV, that are requested
to produce the synchrotron radiation observed in the radio band
\cite{brunetti01,petrosian01,fujita03,cassanobrunetti05,bl07,bl11acc,beresnyak13,donnert13,miniati15}.
This scenario allows to explain the very extended and
diffuse nature
of the observed emission of {\it radio halos} and the tight connection
between {\it radio halos} and dynamics of the hosting clusters.
On the other hand crucial ingredients in this scenario are poorly
known. Primarily the challenge is to understand the chain of mechanisms
that transport energy from large scales to collisionless small-scales
in the ICM.

\noindent
In this paper we focus on CRe acceleration by compressive turbulence
driven at large scales in the ICM. Specifically
we explore the changes in the efficiency
of acceleration that are induced by different assumptions on 
the turbulent spectrum and ICM microphysics.

\section{A brief overview of turbulence in the ICM}

In this Section we briefly review the current view of 
the properties of turbulence in galaxy clusters.
Galaxy clusters contain many potential sources of turbulence
\cite{brunettijones14,bruggenvazza15}, 
the most important for large-scale turbulence
are mergers between clusters.
Mergers deeply stir and rearrange the cluster structure
generating turbulence through sloshing of cluster cores,  
shearing instabilities in the ICM and via the
complex patterns of interacting shocks that form during mergers 
and structure formation more generally.
Such a complex ensemble of mechanisms should drive 
both compressive and incompressive turbulence, as also
supported by the analysis of numerical (fluid)
simulations\cite{zuhone13,beresnyak13,miniati14}.

Large-scale motions that are driven during cluster-cluster
mergers and dark matter sub-halo motions are expected on scales 
comparable to cluster cores scales, $L_o \sim 100-400$ kpc, 
and might have typical velocities $\delta V \sim 300-700 $ km s$^{-1}$
\cite{bruggenvazza15,cassanobrunetti05,subramanian06,vazza11,iapichino11,miniati14}. 
These motions are subsonic, typically with $M_o = \delta V/c_s \approx
0.2-0.5$, but super-Alfv\'enic, with $M_A = \delta V/V_A \approx 2-8$.
An open question is whether these motions generate an efficient turbulent
cascade at smaller scales or if they are dissipated at larger
scales.
The Coulomb mfp of particles in the ICM is very large, $l_C \approx
10$ kpc, which in turns implies a moderate value of the 
{\it Reynolds} number in the ICM, ${\cal R}e \sim 100$, for typical
parameters.
Such a moderate {\it Reynolds} number
would not guarantee {\it per se} an efficient turbulent cascade.
However the ICM is a magnetised and {\it weakly collisional} high-beta
medium, that is notoriously unstable to several instabilities
\cite{levinson92,pistinner96,s10,bl11mfp,yanlazarian11,santoslima14},
implying that collisionless effects govern microphysics and make the
medium more turbulent.
The combination of these facts suggests that the ICM is turbulent.
We note that 
this conclusion is also supported by similarities with the IPM, that 
is also {\it weakly collisional}, magnetised and (moderately) high-beta
plasma. In fact the {\it Reynolds} 
number (considering Coulomb collisions) of the IPM would not be large, but the IPM is 
{\it observed} to be turbulent \cite{alexandrova14}.

At large scales, where the bulk of turbulence is generated by
cluster dynamics in the ICM, 
turbulence is hydrodynamic with kinetic energy
of motions being in excess of magnetic energy. 
At scales smaller than the MHD-scale, $l_A = L_o M_A^{-3}$ (using
Kolmogorov scaling for hydro-turbulence), 
turbulence becomes MHD provided that the ICM
behaves a fluid and that collisionless effects play a role only
at smaller scales. MHD turbulence can be described by Goldreuch-Sridhar
model with Alfv\'en and slow modes developing anisotropic 
spectrum\cite{gs95,cholaz03}. 
On one hand we can thought that solenoidal and compressive turbulence
are generated at large scales and cascade at smaller scales, at the
same time however we should expect that turbulence can also be generated 
directly at 
smaller scales in response of plasma/kinetic instabilities and coherent
wave phenomena driven by the cascade of strong MHD turbulence
\cite{pistinner96,yanlazarian11,wentzel74,w99}; the energy
associated with small-scale turbulence is however subdominant.
Both large-scale motions (and their cascading at smaller scales) and the
variety of waves excited at small scales should play crucial roles in
governing the
micro-physics of the ICM through the scattering of particles and the
perturbation and amplification of the magnetic field.
However, the efficiency of the transport of turbulent energy from
large to small scales and the efficiency of generation of waves at
smaller scales are still open issues.

The nonlinear interplay between particles and turbulent 
waves/modes induces a stochastic process that drains energy from plasma 
turbulence to particles \cite{skilling75,melrose,schlickeiser}.
Turbulent acceleration is invoked for the origin of {\it giant radio
halos}.
Acceleration of CRs directly from the thermal pool to relativistic
energies by MHD turbulence in the ICM is very inefficient and faces
serious problems due to associated energy arguments \cite{pe08}.
Consequently, turbulent acceleration in the ICM is
rather a matter of reacceleration of pre-existing (seed) CRs rather
than {\it ab initio} acceleration of CRs. This poses the problem of the
origin of the seed particles, a problem that is currently subject of active
discussion\cite{bl11acc,pinzke15}, but that will not be addressed in
this paper.

\noindent
Presumably the many types of waves generated/excited in the ICM, both at
large and very small scales, jointly contribute to the
scattering process and (re)acceleration of CRs.
In the last years much attention has been devoted to CR reacceleration
due to compressible turbulence that is driven at large scales
in the ICM from cluster mergers and that cascades to smaller
scales\cite{bl07,miniati15}. This is the simplest scenario that can be 
thought, nevertheless it naturally predicts   
a direct connection between cluster mergers and {\it radio halos}.
In fact several studies suggest that turbulent reacceleration
by compressive modes provides a plausible explanation for
{\it radio
halos}\cite{bl07,bl11acc,donnert13,beresnyak13,brunettijones14,pinzke15}, although several
ingredients of the physics of this mechanim are still poorly known.

\noindent
In the following we will discuss 
Transit-Time-Damping (TTD) resonance (Sect.3) and non-resonant
acceleration by turbulent compressions (Sect.4), exploring the 
effects induced on the efficiency of these mechanisms by different assumptions.

\section{Transit Time Acceleration mechanism: turbulent
spectrum and mfp}

Compressible component of the magnetic field of compressible modes
(i.e. the component along $B_o$ in the case
of oblique propagation) can interact with particles through
TTD resonance \cite{fisk76,schlickeiser98}. The condition for resonance is :

\begin{equation}
\omega - k_{\Vert} v_{\Vert} =0
\label{resonance}
\end{equation}

\noindent
where 
$k_{\Vert}$ and $v_{\Vert}$ are the components
of the waves wavenumber and particle velocity parallel to the
magnetic field; $\omega= c_s k$ for fast modes (magnetosonic waves) in
the ICM.
This interaction is essentially a coupling
between the magnetic moment of particles and the (parallel)
magnetic field gradients.
This interaction, in combination with 
{\it additional/external} sources of pitch-angle scattering/isotropization 
during
acceleration\cite{taaheri85,paesoldbenz99,bl07}\footnote{indeed the 
$n=0$ resonance changes only the component
of particle momentum parallel to the seed magnetic field that would
increase the degree of anisotropy of particle distribution and decrease
the acceleration efficiency with time}, is considered as a 
fundamental way to accelerate particles in different astrophysical 
environments, including the ICM.

\noindent 
Stochastic acceleration can be described as a diffusion in the momentum
space of particles. For TTD the expression of diffusion coefficient,
assuming quasi-isotropic turbulent cascade and
high-beta plasma (conditions
suitable for the ICM), is given in \cite{bl07}:

\begin{equation}
D_{pp}(p)=
{{\pi^2}\over{2 c}}
{{p^2 c_s^2}\over{B_o^2}}
\int_0^{1} d\mu {{ 1 - \mu^2 }\over{\mu}}
{\cal H}\left(1 - {{c_s}\over{c \mu}} \right)
\left[
1 - ( {{c_s}\over{c \mu}} )^2 \right]
\int_{k_o}^{k_{cut}} dk W_B(k) k
\label{dppTTD}
\end{equation}

\noindent
where $\mu$= cosine of pitch angle, 
${\cal H}(x)$ is the Heaviside step function (1 for $x>0$,
and 0 otherwise), $W_B$ is the energy spectrum of magnetic field
fluctuations and $k_o$ and $k_{cut}$ are the injection and cut-off
scales (wavenumber) of the fluctuations.

\noindent
The most important ingredients in Eq.\ref{dppTTD}
are the energy density of electromagnetic fluctuations and their
spectrum, and the cut-off scale
of the turbulent (magnetic fluctuations) spectrum $k_{cut}$.
In the simple scenario adopted in this paper,
where turbulence is generated at large scale
and cascade at smaller scales, the minimum scale $k_{cut}$ is the scale
where the mechanisms of
damping of turbulence become faster than turbulent cascade.
During acceleration energy goes into
particles increasing the damping of turbulence due to particles
themselves (the cut-off scale $1/k_{cut}$ increases)
and reducing the acceleration efficiency.
Constraining $k_{cut}$ is thus critical to obtain meaningful 
estimates of the acceleration 
efficiency.

\noindent
To do that we need to calculate the damping of turbulence in the ICM
and the efficiency of turbulent cascade. Following the motivations
given in \cite{bl07} we assume that collisionless dampings with
particles are the dominant ones in the ICM.
The damping is obtained assuming quasi-linear-theory
\cite{melrose,schlickeiser}:

\begin{equation}
\Gamma = - {\it i} \Big(
{{ E_i^* K^a_{ij} E_j }\over{ 16 \pi W }}
\Big)_{\omega_i =0}\omega_r
\label{damping_start_noA}
\end{equation}

\noindent
where $K_{ij}^a$ is the anti-Hermitian part of the plasma dielectric
tensor \cite{melrose}, $W$ is the total energy in the modes, 
$E_i$ is the electric field (fluctuations) 
and $\omega_r$ is the real part of the mode frequency.

\noindent
TTD determines the strongest collisionless
interaction between particles and compressive fast modes in the ICM. 
The collisionless damping with thermal electrons and protons is
\cite{bl07}:

\begin{eqnarray}
\Gamma_{e/p}(k,\theta) =
\sqrt{ {{\pi}\over 8} }
{{ |B_k |^2}\over{W(k,\theta)}}
{\cal H}
\Big(1- {{c_{\rm s}}\over{c}} {{k}\over{ |k_{\Vert}| }} \Big)
{{ c_{\rm s}^2}\over{B_o^2}}
\left( {{k}\over{|k_{\Vert}|}} \right)
\left( {{k_{\perp}}\over{k}} \right)^2 \times \nonumber\\
{{ \left( m_{e/p} k_B T \right)^{1/2}}\over{
1 - ( {{c_{\rm s} k}\over{c k_{\Vert} }} )^2 }} N_{e/p}
\exp \Big\{ - {{ m_{e/p} c_{\rm s}^2 }\over
{2 k_B T}} {{\left( {{k / k_{\Vert}}} \right)^2 }\over{
1 - ( {{c_{\rm s} k}\over{c k_{\Vert} }} )^2 }}
\Big\} k 
\label{damping_th_LFM_noA}
\end{eqnarray}

\noindent
where the ratio of magnetic field fluctuations and
total energy density is calculated in the collisionless regime
in \cite{bl07}, $\beta_{pl}=2 c_s^2/V_A^2$, and 
$\langle |B_k|^2/W \rangle \approx 16 \pi/\beta_{pl}$, $<..>$ is
the average over pitch-angles.

\noindent
The other source of damping of fast modes in the ICM 
is due to TTD interaction with CRs \cite{bl07}:

\begin{eqnarray}
\Gamma_{e/p}(k,\theta)=
- {{\pi^2}\over{8}}
{{
| B_k |^2 }\over{W(k,\theta)}}
\left(
{{k_{\perp}}\over{k}}
\right)^2
\left(
{{k}\over{|k_{\Vert}|}}
\right)
{\cal H}
\Big(1- {{c_{\rm s}}\over{c}} {{k}\over{|k_{\Vert}|}} \Big)
{{ N_{e^{\pm}/p} \,\, c_{\rm s}^2}\over
{B_o^2}} \,\, k \times \nonumber\\
\big(
1 - ( {{c_{\rm s} k}\over{c k_{\Vert} }} )^2 \big)^2
\int^{\infty}
p^4 dp
\left( {{\partial \hat{f}(p)
}\over{\partial p}}\right)_{e/p}
\label{damping_ur_noA}
\end{eqnarray}

\noindent
where $N_{e^{\pm}/p} \hat{f}(p)_{e/p}$ is the distribution function of
CRe/p in the momentum space.
\noindent
The second element that is necessary to constrain $k_{cut}$ is 
the time-scale of turbulent cascade.
We shall use MHD as a guide and derive
the cascading of isotropic fast modes using the Kraichnan treatment.
In this case the wave-wave diffusion coefficient in k-space is
\cite{zm90}
$D_{kk} \sim k^4 W(k)/(\rho c_s)$ and the resulting cascading time
is :

\begin{equation}
\tau_{kk} \approx {{k^3}\over{(\partial / \partial k) (k^2 D_{kk})}}
\sim {2 \over 9} {{c_s}\over{\delta V^2}} \left( k k_o \right)^{-1/2}
\label{taukk}
\end{equation}

\noindent
In Eq.\ref{taukk} 
and in the following we assume $W(k_o)k_o \sim \delta V^2$, $\delta V$
is the velocity of large-scale eddies.

\noindent
Having derived damping and cascading coefficients, the cut-off scale 
is obtained requiring $\tau_{kk} \sim \Gamma^{-1}$ :

\begin{equation}
k_{cut, K} \simeq {{81}\over{4}}
\left(
{{\delta V^2}\over{c_s}}
\right)^2
{{ k_o }\over{ ( \sum_{\alpha} \langle 
\Gamma_{\alpha} \rangle k^{-1} )^2}}
\label{kcutk}
\end{equation}

\begin{figure}
\begin{center}
\includegraphics[width=0.85\textwidth]{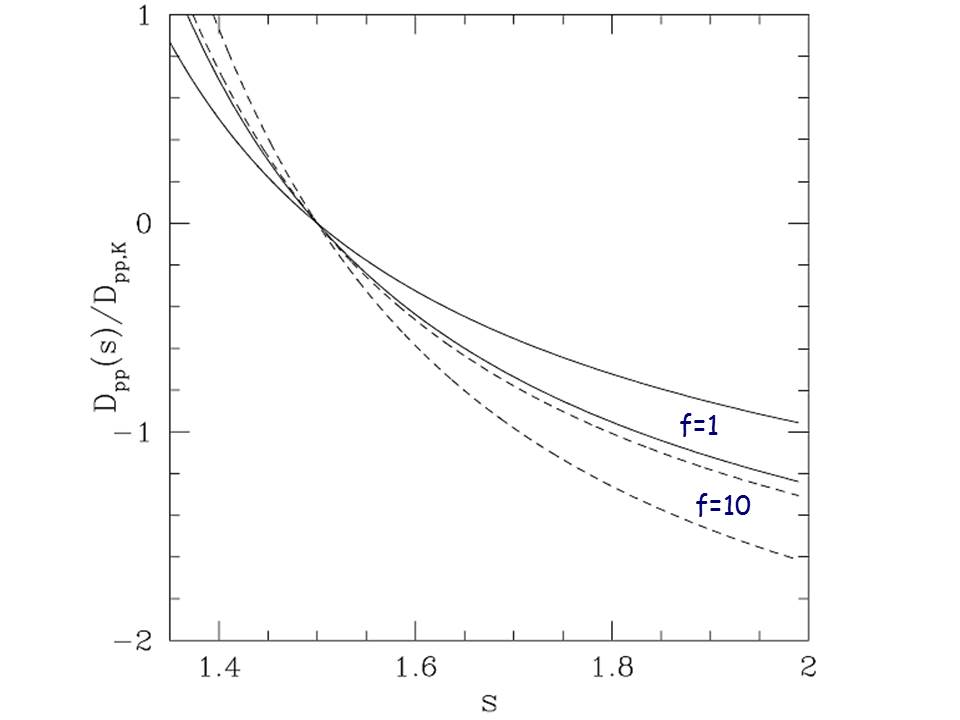}
\end{center}
\caption{The ratio between acceleration rate assuming 
$W_B \propto k^{-s}$ and that using a Kraichnan spectrum.
Solid lines assume collisionless interaction with both thermal particles
and CRs (case (i) in Sect.3), dashed lines assume an 
increased collisionality, with $k_{cut}=f k_{cut,K}$
(to mimic case (ii) in Sect.3).
We adopt $\delta V = 500$ (lower) and
800 km/s (upper lines), $L_o = 300$ kpc, $c_s=1500$ km/s, and
$V_A = 300$ km/s.}
\end{figure}

\noindent
where $<..>$ marks pitch-angle averaging.

\noindent
Given the uncertainties in the ICM microphysics, we consider two
scenarios:

(i) assume that the interaction between turbulent modes and
both thermal and CRs is fully collisionless.
This happens when particles collision frequency in the ICM is $\omega_{ii}
< \omega = k c_{s}$; for example this is the case where ion-ion collisions
in the thermal ICM are due to Coulomb collisions.
Under this condition the damping of compressive modes 
is dominated by TTD with thermal particles (Eq.\ref{damping_th_LFM_noA})
\cite{bl07}. For large $\beta_{pl}$ the dominant damping rate is due 
to thermal electrons and can be 
approximated by $\Gamma_e \sim
c_s k \sqrt{3 \pi (m_e/m_p) / 20 \mu^2} \exp( -5 (m_e/m_p)/3 \mu^2)
(1 -\mu^2)$. Combining pitch-angle averaging of this expression with
Eq.\ref{kcutk} gives $k_{cut,K} \sim 10^4 k_o M_o^4$.

\noindent
Assuming a Kraichnan spectrum of magnetic field fluctuations, $W_B
\propto k^{-3/2}$, the resulting acceleration time-scale 
(from Eq. \ref{dppTTD}) is :

\begin{equation}
\tau_{acc} = {{p^2}\over{4 D_{pp}}} 
\simeq 2.5 \langle {{\beta_{pl} |B_k|^2 }\over{16 \pi W}}
\rangle^{-1} f_x^{-1} ({{M_o}\over{1/2}})^{-4} \left(
{{L_o/300 \, {\rm kpc}}\over{c_s/1500\, {\rm km \,s^{-1}}}}
\right)
\,\, ({\rm Myr})
\label{timeaccC}
\end{equation}

\noindent
where $<..> \sim 1$ independent of scale, $x= c_s/c$ and 

\begin{equation}
f_x = x \left( {{x^4}\over{4}} + x^2
-(1+2x^2)\ln(x) -{5 \over 4} \right) \simeq 0.02
\end{equation}

\noindent
For typical conditions this scenario predicts acceleration
time-scales $\sim 100$ Myr that are sufficient to explain {\it radio
halos} and is commonly adopted to calculate turbulent 
acceleration in the ICM \cite{bl07,db14,miniati15,pinzke15}. 

(ii) The other possibility is that 
thermal particles do not contribute
very much to the collisionless dampings.
The ICM is a {\it weakly collisional} high-beta plasma that is unstable
to several instabilities (see Sect.2).
Scattering induced by magnetic field perturbations driven by
instabilities may increase the collision frequencies in the thermal 
plasma because charged particles
can be randomized if they interact with the perturbed magnetic
field. This process can
be viewed as the collective interaction of an individual ion with the
rest of the plasma.
Under this condition one may thought that the interaction between
modes and thermal particles behave {\it collisional}.
\footnote{It should also be mentioned that scatterings induced by instabilities
decrease the effective mfp of thermal particles decreasing the
effective viscosity in the ICM and increasing the effective
Reynolds number of the ICM.}
In this case the dominant collisionless damping in the ICM is 
due to the CRs \cite{bl11mfp}, and combining Eqs.\ref{damping_ur_noA} 
and \ref{kcutk} :

\begin{equation}
k_{cut,K} \simeq
\left( {{18}\over{\pi^2}} \right)^2
M_o^4 f_x^{-2} {{ (\rho c_s^2)^2 }\over{
\left[ 
c \int p^4 dp {{\partial f}\over{\partial p}} \right]^2
}} k_o
\label{kcutCR}
\end{equation}

\noindent
Under typical conditions in the ICM this is $k_{cut,K} \sim 1000 \,
k_o M_o^4 (\epsilon_{ICM}/\epsilon_{CR})^2$, where the ratio of
thermal and CR energy densities is $\epsilon_{ICM}/\epsilon_{CR} \sim
100$. It implies that 
the turbulent cascade
reaches scales that are much smaller than those in the case (i) and 
consequently the acceleration rate is faster (Eqs.\ref{dppTTD} and
\ref{kcutCR}) :

\begin{equation}
\tau_{acc} = {{p^2}\over{4 D_{pp}}}
\simeq 6 \langle {{\beta_{pl} |B_k|^2 }\over{16 \pi W}}
\rangle^{-1} ({{M_o}\over{1/2}})^{-4} \left(
{{L_o/300 \, {\rm kpc}}\over{c_s/1500\, {\rm km \,s^{-1}}}}
\right)
(\left[
{{c \int p^4 dp {{\partial f}\over{\partial p}} }\over
{\rho c_s^2 }} \right] / 25)
\,\, ({\rm Myr})
\label{timeaccCR}
\end{equation}

\noindent
The acceleration efficiency is inversely proportional to the
energy density of CRs. In this scenario 
CRs drain efficiently energy from the
turbulent cascade, however as the CRs energy density increases the
acceleration efficiency gets reduced.
From the practical point of view this is simply 
because CRs can get a constant energy flux from turbulence implying that
the effect on their spectrum is smaller for increasing values of the
CRs energy density. As a consequence of this {\it back reaction} fast
acceleration cannot be maintained for long time. 
Assuming typical conditions in the ICM, the value
of the acceleration time averaged on a time-period of few 100 Myr is
generally found $\tau_{acc}\sim$ 10 Myrs \cite{bl11mfp}.
This is up to 10 times more efficient than in the case (i)
resulting in harder spectra of both CRe and synchrotron
radiation\cite{bl11mfp, miniati15}.

All the 
results discussed above are based on Kraichnan spectrum of fast modes,
$W_B(k) \propto k^{-3/2}$. 
On the other hand part of the energy of 
the MHD cascade of fast modes may be dissipated into weak shocks
producing a steeper spectrum\cite{porter15}.
In general if the spectrum of compressive turbulence gets steeper
the turbulent acceleration rate decreases.
Assuming a spectrum $W_B(k) \propto k^{-s}$ in Eq.\ref{dppTTD},
the acceleration rate will be changed with respect to that 
evaluated using a Kraichnan spectrum by a factor\footnote{in the
case of Burgers spectum, $s=2$, the factor is
$\sim (1/2) \ln(k_{cut,2}/k_o) \sqrt{k_o/k_{cut,K}}$}
$\sim {{1/2} \over{2-s}} k_o^{s-1} k_{cut,s}^{2-s}/\sqrt{k_o k_{cut,K}}$,
where $k_{cut,s}$ is the cut-off in the turbulent spectrum
assuming a slope $s$. 

\noindent
In order to evaluate this $k_{cut,s}$ we follow a
simple procedure. We still assume the Kraichnan diffusion coefficient 
$D_{kk} \sim k^4 W(k)/\rho c_s$ to model wave-wave coupling at scale
$k$, but we use a spectrum $W(k) \propto k^{-s}$.
The resulting cascading time scale of fast modes is :

\begin{equation}
\tau_{kk} \approx {1 \over{5-s}} {{c_s}\over{\delta V^2}}
{{k^{s-2}}\over{k_o^{s-1}}}
= {{7/2}\over{5-s}} \tau_{kk}(s={3\over 2}) \left(
{{k}\over{k_o}} \right)^{s-{3 \over 2}}
\label{taukks}
\end{equation}

\noindent
and the cut-off, obtained from the condition
$\tau_{kk} \sim 1/\sum \langle \Gamma \rangle$, is :

\begin{equation}
k_{cut,s} =
\left[ {2 \over 7}
(5-s)
k_{cut,K}^{1/2}
k_o^{s-3/2}
\right]^{1 \over{s-1}}
\label{kcutkcutK}
\end{equation}

\noindent
where $k_{cut,K}$ refers to the cut-off in both Eqs.\ref{kcutk} and
\ref{kcutCR}.

\noindent
Figure 1 shows the effect on the acceleration rate due to 
different assumptions for the turbulent spectrum. 
The acceleration rate decreases with
increasing slope. For Burgers spectra ($s=$2) the acceleration is 10
times less efficient that that in the Kraichnan case.
This has important consequences, for example in the collisional case
(case i) steep spectra of magnetic field fluctuations produce
acceleration rate via TTD that are generally too small to explain {\it
radio halos} \cite{miniati15}. 

\section{Stochastic Acceleration by large-scale compression: turbulent
spectrum and mfp of relativistic particles}

\begin{figure}
\begin{center}
\includegraphics[width=0.494\textwidth]{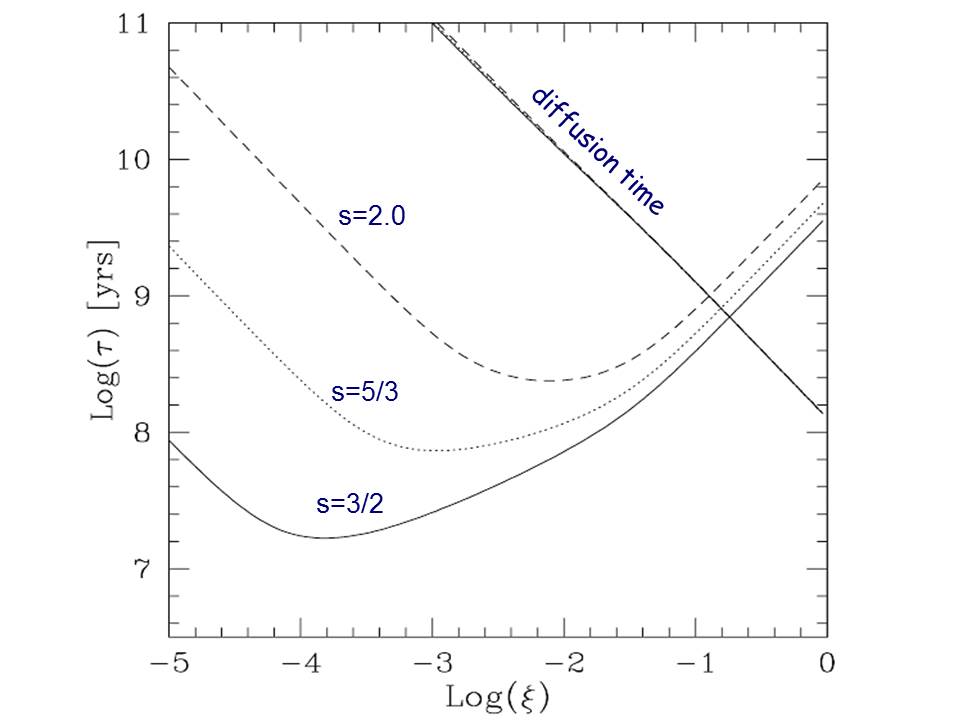}
\includegraphics[width=0.494\textwidth]{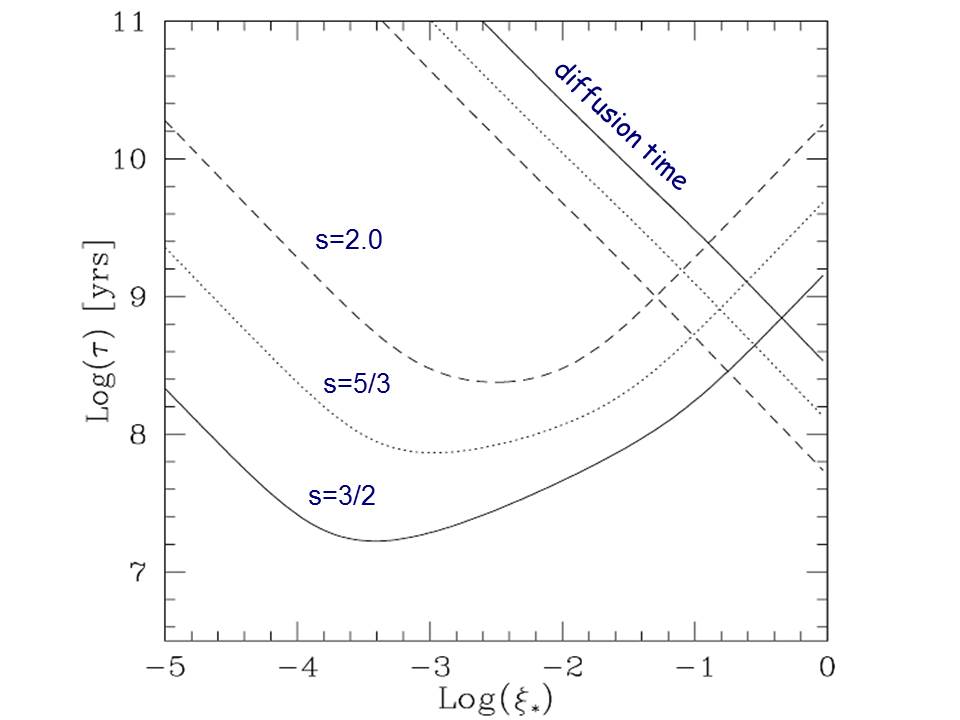}
\end{center}
\caption{
Acceleration time and (Mpc-scale) diffusion time due to non-resonant
acceleration as a function of $\xi =
\lambda_{mfp}/l_{\lambda}$. Left panel is for $l_{\lambda} =
M_o^{-3} L_o$ and right panel is for $l_{\lambda} = {\rm max}
\{M_o^{-4} L_o, \, 2\pi /k_{cut} \}$ (see text).
The collisionless scenario (case (i),
Sect. 3) is adopted to calculate the turbulent cut-off scale.
In the calculations we assume $\delta V$=750 km/s, and $s=$3/2, 5/3, 2.0,
other parameters being equal to Fig.1.}
\end{figure}

Relativistic particles diffusing through large-scale compressible
turbulence experience a statistical acceleration effect
\cite{p88,cho06,bl07}.
In the case of subsonic turbulence, $\delta V^2 << c_s^2$, and provided
that turbulence has correlation scales much longer than the particles
mfp, the diffusion coefficient in the particle momentum space is :

\begin{equation}
D_{pp} =
{2 \over 9}
p^2 D \int_{k}
{{ dy y^2 {\cal K}(y) }\over
{c_s^2 +y^2 D^2 }}
\label{dpp_compression_1}
\end{equation}

\noindent
where $D$ is the particle spatial diffusion coefficient and $\cal{K}$ is
the kinetic spectrum of turbulence, $k_o {\cal K}(k_o) \sim \delta V^2/2$.

\noindent
The spatial diffusion coefficient of CR in the ICM is unknown and we
shall consider it as a free parameter.
We assume $D= {1\over 3} c \lambda_{mfp}$ with a CR mfp 
$\lambda_{mfp} = \xi l_{\lambda}$, where 
$l_{\lambda}$ is the minimum scale of magnetic field reversal
in the ICM. 
Under these assumptions Eq.\ref{dpp_compression_1} reads :

\begin{equation}
D_{pp}=
{{p^3}\over 3}
\left(
{{\delta V^2 }\over{c l_{\lambda}}} \right)
\xi
\int_1^{x_c}
{{ dx x^{2-s} }\over
{( {{3 c_s L_o}\over{2 \pi c l_{\lambda}}} )^2 +
x^2 \xi^2}}
\label{dppcompressionpara}
\end{equation}

\noindent
where $x_c = k_{cut}/k_o$.
The acceleration rate depends on the turbulent energy and spectrum, but
also on the CRs mfp.
This mechanims is characterised by two regimes : {\it fast diffusion}, for
$\xi^2 >> (3 c_s L_o)^2/(2\pi c l_{\lambda})^2$, in which case particles leave
the eddies before they turnover and the acceleration is dominated by the
largest eddies, and {\it slow diffusion}, for 
$x_c^2 \xi^2 << (3 c_s L_o)^2/(2\pi c l_{\lambda})^2$, in which case the
acceleration is mainly dominated by the smallest eddies.
In the two regimes the dependencies of the acceleration rate on the
physical parameters are different: $D_{pp} \propto p^2 M_o^2 D k_o^{s-1}
k_{cut}^{3-s}$ in the {\it slow} regime and 
$D_{pp} \propto p^2 M_o^2 c_s^2/D$ in the {\it fast} regime (see also
\cite{cho06}).

\noindent
The total (turbulent advection and diffusion) spatial
diffusion coefficient of particles diffusing and interacting
with compressible/acoustic turbulence is \cite{p88,cho06,bl07} :

\begin{equation}
D_* =
{1 \over 3}
l_{\lambda} c \xi
\left[
1 +
{3 \over{2 \pi^2}}
\left( {{L_o \delta V}\over{l_{\lambda} c}} \right)^2
\int_1^{x_c}
{{ dx x^{-s} }\over
{( {{3 c_s L_o}\over{2 \pi c l_{\lambda}}} )^2 +
x^2 \xi^2}}
\right]
\end{equation}

\noindent
the time-scale necessary to diffuse on scale $L_*$ is $\tau_{diff}
\simeq L_*^2/4 D_*$.

\begin{figure}
\begin{center}
\includegraphics[width=0.494\textwidth]{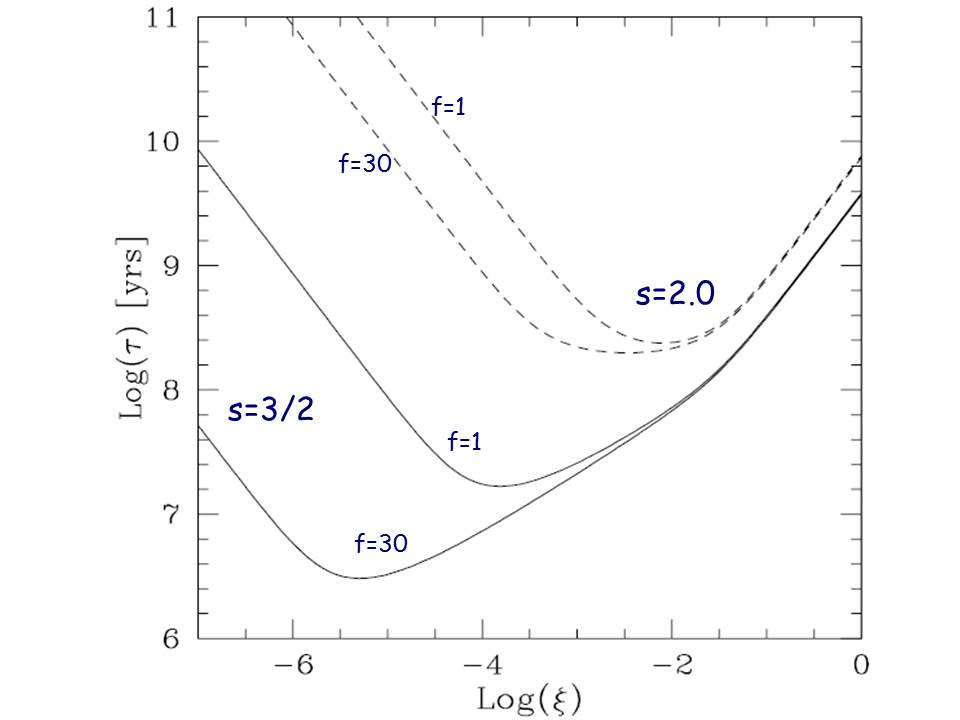}
\includegraphics[width=0.494\textwidth]{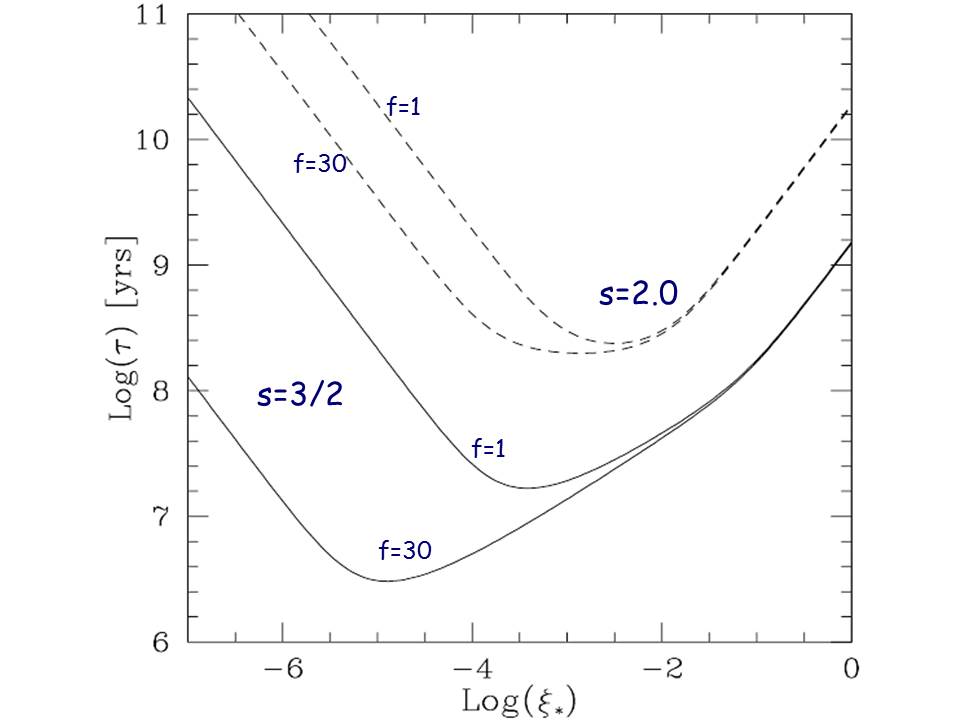}
\end{center}
\caption{
Acceleration time due to non-resonant
acceleration as a function of $\xi =
\lambda_{mfp}/l_{\lambda}$. Left panel is for $l_{\lambda} =
M_o^{-3} L_o$ and right panel is for $l_{\lambda} = {\rm max}
\{M_o^{-4} L_o, \, 2\pi /k_{cut} \}$ (see text).
Turbulent cut-off scale is calculated both in the 
collisionless scenario ($k_{cut,s}$ Eqs.7 and 10, 
case (i), Sect. 3) and assuming $k_{cut} = f k_{cut,s}$ 
(to mimic case (ii) in Sect.3).
We show the case $s=$3/2 (solid) and 2.0 (dashed), 
other parameters being equal to Fig. 2.}
\end{figure}

\noindent
In order to quantify the acceleration rate we should estimate
$l_{\lambda}$. This reversal depends on turbulent properties.
We estimate $l_{\lambda}$ following two hypothesis : 
(a) we assume that both solenoidal and compressible
turbulence are generated in the ICM at scale $L_o$ with
similar energies, $\delta V_s^2 \approx \delta V^2$, and
use solenoidal turbulence to estimate $l_{\lambda}= l_{As}$,
where $l_{As} = L_o M_A^{-3}$ is the MHD scale assuming Kolmogorov
spectrum of the (super-Alfv\'enic) solenoidal turbulence; (b) we assume
that turbulence in the ICM
is only compressible, in this case $l_{\lambda} =
{\rm max}\left\{l_A, 2\pi/k_{cut} \right\}$ where $l_{A} = L_o M_A^{-4}$
is the MHD scale (for Kraichnan turbulence) and $k_{cut}$ is given in
Sect.3.

\noindent
Fig.2 shows the acceleration rate vs $\xi$ assuming different slopes
of the kinetic turbulent spectrum. The cut-off scale is derived
according to Sect.3, case (i).
Although $\xi$ is a free parameter, we note that
mfp$>> r_L$ implies $\xi > 10^{-6}$
for multi-GeV -- TeV particles in the ICM.
At the same time the upper bound of $\xi$ is set by the condition
that CRe must be confined in Mpc-volumes for $\geq$Gyrs in order to
generate {\it giant radio halos}, this typically requires $\xi < 0.1$.
For $\delta V \sim 700-800$ km/s, we conclude that the acceleration 
time ranges between $10^7-{\rm few}\times 10^9$ yrs 
depending on the slope of the
turbulent spectrum and on the mfp of CRs.

\noindent
In Fig.3 we show the effect of increasing plasma collisionality in the
ICM (i.e. case (ii) in Sect.3) by assuming a cut-off $k_{cut}=f k_{cut,s}$.
This increases the acceleration rate in the branch of {\it slow}
diffusion regime. The effect is stronger for flatter spectra of the
turbulence, for instance $f=30$ implies a boost of the acceleration 
efficiency by more than 2 orders of magnitude in the Kraichnan case.

\section{Discussion and conclusions}

A popular scenario that is
adopted to explain {\it giant radio halos} is based on
turbulent reacceleration, with turbulence generated during cluster-cluster
mergers.
If true this scenario implies that a hierarchy of complex mechanisms
drain a fraction of the energy of the large-scale
motions that are generated by the process of cluster formation into electromagnetic 
fluctuations and collisionless mechanisms of particle acceleration at 
much smaller scales.
It has been realised that the existence of these complex collisionless
mechanisms, opens new
prospects to understand the micro-physics of the ICM
\cite{beresnyak13,brunettijones14,miniati15}.

In this paper we focused on the problem of stochastic acceleration of
CRs by compressible turbulence in the ICM.
The efficiency of acceleration depends on the particles mfp and on
the spectrum of 
compressible turbulent motions, in particular on that of the
electromagnetic fluctuations. 
The extent and the shape of this spectrum, in turn depend on the processes of
plasma damping and on the way turbulence is generated and transported
at smaller scales. 
We have explored this subject by 
considering two mechanisms that are commonly adopted to explain
{\it giant radio halos}, TTD due to fast modes (Sect. 3) and
non-resonant acceleration due to turbulent compressions (Sect. 4).

\noindent
In the case of TTD we analyzed two extreme situations that 
differs in the efficiency of collision frequencies between 
thermal particles in the ICM, 
in particular whether collisions occur via Coulomb scattering or
mainly via collective processes induced by plasma instabilities.
In the latter case collisionless damping of compressive turbulence
is dominated by CRs and the acceleration
is more efficient than in the other case.
We also discussed the changes in the acceleration rate that are induced 
by different slopes of the turbulent spectrum.

\noindent
In the case of non-resonant acceleration we explored 
the combined effect induced
on the acceleration rate by different assumptions for
the turbulent spectra and mfp of CRe.
The latter parameter is very uncertain but plays a crucial role
as it determine the regime of diffusion of CRe in the 
turbulent field and consequently has the potential to strongly change the
acceleration efficiency.

\noindent
In conclusion we have shown that the uncertainties about the ICM
microphysics induce substantial variations in the acceleration efficiency of both
mechanisms.
On the other hand however this also implies that radio halos and 
the non-thermal properties of galaxy clusters are effective 
probes of the complex microphysics of the ICM.

In this paper we do not investigate 
reacceleration by solenoidal/incompressive turbulence.
Several works attempted to model this
situation to explain radio halos\cite{ohno02,fujita03,brunetti04}.
We note however that these calculations
did not take into account the scale-dependent anisotropies in Alfv\'enic
turbulence, and consequently addressing the role of Alfv\'enic
turbulence in the acceleration process in the ICM requires further
investigations.

At this point we believe that future advances in the field will derive 
from studies that aim at addressing the generation of plasma instabilities 
in the ICM and from attempts to model self-consistently the way these
instabilities generate small-scale fluctuations, 
and affect particles mfp and acceleration rates.
The importance of this step has been clearly highlighted in Sect.3 and 4 
where we have compared acceleration rates obtained 
in the collisional and collisionless cases. Situation may be even more
complicated because, similarly to the IPM, it can be thought that the 
collisional properties of the ICM may evolve with time and space.
Finally we believe that magnetic reconnection in the ICM and its interplay with turbulence
is another piece of this complex puzzle and it may play a role in the
acceleration and reacceleration of CRe.

\section*{Acknowledgments}

The author acknowledge the two referees for useful comments.
The author acknowledges support from the Alexander von Humboldt
Foundation and from PRIN-INAF 2014.
During the preparation of the manuscript the author has been hosted as
Humboldt awardee at the Dep. of Theoretical Physics IV of  
the Bochum University, and acknowledges warm hospitality 
by Prof. R. Schlickeiser and his group.

\end{document}